\documentclass{ws-mpla}

\def\de{{\rm d}}
\def\nn{\nonumber \\}

\begin{document}

\markboth{Marcin Badziak}
{Yukawa Unification in SUSY SO(10) in Light of the LHC Higgs Data}

\catchline{}{}{}{}{}

\title{YUKAWA UNIFICATION IN SUSY SO(10) \\
IN LIGHT OF THE LHC HIGGS DATA
}

\author{\footnotesize MARCIN BADZIAK}

\address{Department of Applied Mathematics and Theoretical Physics, 
Centre for Mathematical Sciences, University of Cambridge, 
Wilberforce Road, Cambridge CB3 0WA, United Kingdom \\
and\\
Cavendish Laboratory, University of Cambridge, \\ J.J. Thomson Avenue,  Cambridge CB3 0HE, United Kingdom
\\
M.Badziak@damtp.cam.ac.uk}

\maketitle


\begin{abstract}
The status of top-bottom-tau Yukawa coupling unification in supersymmetric SO(10) models is reviewed with a particular emphasis on the implications of the Higgs boson mass in the vicinity of 125 GeV, as suggested by the LHC Higgs data. In addition, the recently proposed model with negative $\mu$, D-term splitting of the soft scalar masses and non-universal gaugino masses generated by a non-zero F-term in a 24-dimensional representation of SU(5) $\subset$ SO(10) is re-analysed in the context of the 125 GeV Higgs. The condition of top-bottom-tau Yukawa unification together with the Higgs mass of about 125 GeV impose strong lower mass limits on SUSY particles. Nevertheless, some of the MSSM particles may be within the reach of the LHC. In the case of models with positive $\mu$ this is the gluino. While in the case of negative $\mu$ these are the pseudoscalar Higgs, the lighter sbottom (sometimes strongly degenerate with the LSP leading to sbottom coannihilations), the right-handed down squark and the gluino.

\end{abstract}

\ccode{PACS Nos.: 12.10.Kt, 12.60.Jv, 14.80.Da, 14.80.Ly}

\section{Introduction}	

The apparent unification of the gauge couplings in the Minimal Supersymmetric Standard Model (MSSM) is often regarded as an indirect evidence for supersymmetry (SUSY). SO(10) is arguably the best candidate for the unified gauge group.
In the minimal version of SUSY SO(10) Grand Unified Theory (GUT) the Yukawa couplings of top, bottom and tau unify at the GUT scale. The Yukawa couplings in the MSSM are quite sensitive to the low scale SUSY threshold corrections. From the bottom-up perspective this means that the values of the Yukawa couplings at the GUT scale depend on the low energy MSSM spectrum. Therefore, the condition of top-bottom-tau Yukawa unification constrains the MSSM spectrum making the SO(10) models very predictive.

Early studies of the Yukawa unification were motivated by the large hierarchy between the top and bottom masses and focused on the unification of the top and bottom Yukawa couplings. It was shown in Ref.~\refcite{mosp_tbuni} that the condition of proper radiative electroweak symmetry breaking (REWSB) imposes important constraints on this scenario and favours the GUT scale values of the universal gaugino mass, $M_{1/2}$, much larger than those of the universal scalar mass, $m_0$. In early 1990s, before the discovery of the top quark, it was pointed out that the assumption of full top-bottom-tau Yukawa unification leads to the prediction for the top quark mass.\cite{tbtau_toppred}  It was then shown that the prediction for the top quark mass suffers from large uncertainties due to the low scale SUSY threshold corrections to the bottom mass which depend on the (unknown) MSSM spectrum.\cite{Hall,Carena} Those correction were proven to be very large in the constrained MSSM (CMSSM).\cite{Carena} The top quark mass measured later at the Tevatron turned out to be in a rather good agreement with the primary prediction for the top quark mass\cite{tbtau_toppred} that neglected the low scale SUSY threshold corrections to the Yukawa couplings. That implied incompatibility of top-bottom-tau Yukawa unification with the assumption of universal soft terms at the GUT scale. Patterns of non-universal scalar masses that suppress the SUSY threshold corrections to the bottom mass and make top-bottom-tau Yukawa unification viable were identified in Ref.~\refcite{Olechowski,Matalliotakis,Murayama}. 

The phenomenological implications of top-bottom-tau Yukawa unification strongly depend on the sign of the Higgs-mixing parameter $\mu$.\footnote{We use the sign convention in which the gluino mass parameter, $M_3$, is positive.} For the universal gaugino masses the negative sign of $\mu$ is disfavoured by the constraints from $(g-2)_{\mu}$ and BR$(b\to s \gamma)$. This is the reason why most of the studies of top-bottom-tau Yukawa unification focused on the case of positive $\mu$, see e.g. Ref.~\refcite{Blazek1,Blazek2,Auto,BaerDM}. 
In the recent years there was a renewal of interest in models with negative $\mu$ because well motivated patterns of non-universal gaugino masses have been identified that ease the tension with $(g-2)_{\mu}$ and BR$(b\to s \gamma)$ constraints.\cite{GoKhRaSh,Gogoladze_HS_negmu,bop}

In this brief review we discuss the current status of the SO(10) models predicting top-bottom-tau Yukawa unification. We give a particular emphasis to the implications of the recent ATLAS\cite{atlas_higgs} and CMS\cite{cms_higgs} Higgs searches which suggest the existence of the SM-like Higgs boson with the mass in the vicinity of 125 GeV. In addition, we perform a dedicated analysis of an impact of the 125 GeV Higgs on the recently proposed model\cite{bop} with negative $\mu$ and non-universal gaugino masses generated by a non-zero F-term in a 24-dimensional representation of SU(5) $\subset$ SO(10).

The article is organised as follows. In section \ref{sec:cond} we discuss some necessary conditions for top-bottom-tau Yukawa unification consistent with the radiative electroweak symmetry breaking. We emphasize the role played by the sign of the Higgs-mixing parameter $\mu$. In section \ref{sec:posmu} we focus on the SO(10) models with positive $\mu$. The case of negative $\mu$ is discussed in section \ref{sec:negmu}. The model with non-universal gaugino masses proposed in Ref.~\refcite{bop} with non-universal gaugino masses is analysed in detail. Most of the results presented in section \ref{sec:negmu} have not been published before. In particular, we present novel Yukawa-unified solutions characterized by the Higgs mass of about 125 GeV and a small mass splitting between the lighter sbottom and the LSP which predict relic abundance of dark matter consistent with the cosmological data.
We give a brief summary in section \ref{sec:sum}.

\section{Conditions for Yukawa unification}
\label{sec:cond}

In contrast to gauge coupling unification which is rather generic prediction of the MSSM, there are many obstacles which often prevent top-bottom-tau Yukawa coupling unification to occur. Most of the problems with Yukawa unification can be traced back to a large value of $\tan\beta\approx m_t/m_b$ which is needed to explain the observed hierarchy between the top and bottom masses. First of all, for large values of $\tan\beta$ proper REWSB encounters serious difficulties. Let us explain this point below. At large $\tan\beta$, proper REWSB implies that parameter $\mu$ is given by:
\begin{equation}
\label{mucond}
 \mu^2\approx -m_{H_u}^2-\frac{M_Z^2}{2} \,, 
\end{equation}
while the mass squared of the CP-odd Higgs is:
\begin{equation}
\label{mAcond}
 m_A^2\approx m_{H_d}^2+m_{H_u}^2+2\mu^2 \,.
\end{equation}
Since the value of $\mu^2$ is fixed by Eq.~(\ref{mucond}), the above condition leads to the following condition for the Higgs mass squared splitting:
\begin{equation}
\label{mHumHdcond}
m_{H_d}^2-m_{H_u}^2\gtrsim M_Z^2+m_A^2 \,.
\end{equation}
The recent Higgs searches at the LHC set very stringent lower limits on $m_A$. Those limits are especially strong at large $\tan\beta$  since the production cross-section of the pseudoscalar Higgs is proportional to $\tan^2\beta$. For values of $\tan\beta$ relevant for top-bottom-tau Yukawa unification i.e. between about 40 and 55, the CP-odd Higgs with mass of order ${\mathcal O}(400-500)$ GeV is excluded at 95\% C.L..\cite{cms_mAbound} We should, however, stress, that the problems with REWSB stemming from the condition (\ref{mHumHdcond}) were identified a long time before the LHC era with a very mild assumption that the pseudoscalar Higgs should be non-tachyonic.\cite{mosp_tbuni}  

The condition (\ref{mHumHdcond}) has to be satisfied at the EW scale so the renormalization group (RG) running of $m_{H_d}^2-m_{H_u}^2$ is crucial for REWSB. An appropriate one-loop renormalization group equation (RGE) is given by:\cite{2loopRGE,BaerRGE}
\begin{equation}
\label{mHdmHuRGE}
 8\pi^2\frac{\de}{\de t}(m_{H_d}^2-m_{H_u}^2)=3h_t^2X_t-3h_b^2X_b-h_{\tau}^2X_{\tau}+h_{\nu}^2X_{\nu}+\frac{3}{5}g_1^2S \,,
\end{equation}
where 
\begin{align}
 &X_t=m_{Q_3}^2+m_{U_3}^2+m_{H_u}^2+A_t^2\,, \\
 &X_b=m_{Q_3}^2+m_{D_3}^2+m_{H_d}^2+A_b^2\,, \\
 &X_{\tau}=m_{L_3}^2+m_{E_3}^2+m_{H_d}^2+A_{\tau}^2\,, \\
 &X_{\nu}=m_{L_3}^2+m_{\tilde{\nu}_{R3}}^2+m_{H_u}^2+A_{\nu}^2\,,  \\
  &S=m_{H_u}^2-m_{H_d}^2+\sum_{i=1}^3[m_{Q_i}^2-2m_{U_i}^2+m_{D_i}^2-m_{L_i}^2+m_{E_i}^2] \,,
\end{align}
$t\equiv \ln(M_{\rm in}/Q)$ and we omitted terms proportional to the first and second generation Yukawa couplings which are negligible. We included the effect of the right-handed neutrino which might be relevant if its mass is substantially smaller than the GUT scale.
Notice that for universal soft terms at the GUT scale, $\frac{\de}{\de t}(m_{H_d}^2-m_{H_u}^2)$ vanishes at the GUT scale if $h_t=h_b=h_{\tau}=h_{\nu_{\tau}}$ as predicted by the minimal SO(10) GUT. Therefore, the sign and the magnitude of $m_{H_d}^2-m_{H_u}^2$ at the EW scale is fully determined by the RG running of the rhs of Eq.~(\ref{mHdmHuRGE}). There are to competing effects on the RG running of $m_{H_d}^2-m_{H_u}^2$. First: different hypercharge assignment of top and bottom results in the positive contribution to $m_{H_d}^2-m_{H_u}^2$. This effect becomes more significant as the gaugino (especially gluino) masses increase. Second: Since the right-handed neutrino decouples (at best) only few orders of magnitude below the GUT scale, the tau Yukawa coupling drives $m_{H_d}^2-m_{H_u}^2$ towards negative values and this effect is more significant for larger scalar masses. In the case of universal soft terms at the GUT scale, the impact of the above two effects on the RG running of $m_{H_d}^2-m_{H_u}^2$ is comparable and results in $m_{H_d}^2-m_{H_u}^2 \approx {\mathcal O}(0.1)M_{1/2}-{\mathcal O}(0.1)m_0$ at the EW scale. This implies that the condition (\ref{mHumHdcond}) requires $M_{1/2}\gtrsim m_0$ and is most likely satisfied if $M_{1/2}\gg m_0$. After taking into account the LHC bounds on the CP-odd Higgs boson mass, this also implies that for the universal soft terms the condition (\ref{mHumHdcond}) can be satisfied only for $M_{1/2}\gtrsim1$ TeV resulting in the gluino which is much too heavy to be discovered at the LHC with $\sqrt{s}=8$ TeV.\footnote{The strong lower bound on the gluino mass may be very relevant for the SO(10) model proposed in Ref.~\refcite{altYuk} which predict that only the top and bottom Yukawa couplings unify at GUT scale. }
 
However, it was found that for the universal soft terms at the GUT scale top-bottom-tau Yukawa unification cannot be realized independently of the value of $M_{1/2}$.\cite{Carena} The reason is following. At large $\tan\beta$ there are sizable SUSY corrections to the bottom mass\cite{Hall,Carena,Pierce} which have big impact on bottom-tau Yukawa unification. The main finite corrections originate from the gluino-sbottom and chargino-stop loops and are given by:\cite{Hall,Carena,Pierce} 
\begin{equation}
\label{mbsusycorr}
\left(\frac{\delta m_b}{m_b}\right)^{\rm finite}
\approx\frac{g_3^2}{6\pi^2}\mu m_{\tilde g}\tan\beta \, 
I(m_{\tilde b_1}^2,m_{\tilde b_2}^2,m_{\tilde g}^2)
+\frac{h_t^2}{16\pi^2}\mu A_t\tan\beta \, 
I(m_{\tilde t_1}^2,m_{\tilde t_2}^2,\mu^2) \,,
\end{equation}
where the loop integral $I(x,y,z)$ (given e.g. in the appendix of Ref.~\refcite{Hall}) scales as the inverse of the mass squared of the heaviest particle propagating in the loop and is well approximated by $a/\max(x,y,z)$ with $a$ between $0.5$ and $1$. In order to allow for bottom-tau Yukawa unification the above correction has to be negative with the magnitude of about 10\% to 20\%.\cite{Wells_yuk} The gluino-sbottom contribution dominates in the vast majority of the parameter space since the numerical coefficient in front of this contribution is much larger than the one in front of the chargino-stop contribution. In particular, the sbottom-gluino contribution dominates for $M_{1/2}\gtrsim m_0$, as required by the condition (\ref{mHumHdcond}). 

Assuming the dominance of the sbottom-gluino contribution in Eq.~(\ref{mbsusycorr}), the finite SUSY threshold correction (\ref{mbsusycorr}) can be conveniently rewritten using the experimental value of the strong coupling constant, the properties of the loop integral $I(x,y,z)$ appearing in Eq.~(\ref{mbsusycorr}) and the value of $\tan\beta\approx 50$ required for top-bottom Yukawa unification:
\begin{equation}
\label{mb_gluinocorr}
\left(\frac{\delta m_b}{m_b}\right)^{\tilde{g}\tilde{b}}
\approx {\mathcal O}(0.5 - 1)\frac{\mu}{m_{\tilde g}}\, 
{\rm min}\left(1,\left(\frac{m_{\tilde g}}{m_{\tilde{b}}}\right)^2\right)
,
\end{equation}
where $m_{\tilde{b}}$ is the mass of the heavier sbottom. The above formula together with the requirement that the finite SUSY correction (\ref{mbsusycorr}) to the bottom mass does not exceed 20\% leads to the upper bound for the Higgs-mixing parameter $|\mu|$:\cite{bop}
\begin{equation}
 \label{mubound}
|\mu|\lesssim 0.4 m_{\tilde g}\approx M_{3}\,,
\end{equation}
where $M_3$ is the gluino mass at the GUT scale and it is also assumed that $m_{\tilde{g}}>m_{\tilde b}$ which is usually the case unless $M_{1/2}\ll m_0$.
For universal soft terms at the GUT scale and $M_{1/2}\gtrsim m_0$, $|\mu|>M_3$ so the condition (\ref{mubound}) is violated and the SUSY correction to the bottom mass is too large to allow for top-bottom-tau Yukawa unification. Therefore, the conditions (\ref{mHumHdcond}) and (\ref{mubound}) cannot be simultaneously fulfilled unless some non-universalities of soft terms at the GUT scale are introduced.

Patterns of non-universal scalar masses that may allow for top-bottom-tau Yukawa unification consistent with REWSB under assumption of universal gaugino masses were investigated in Ref.~\refcite{Olechowski}. The necessary condition for top-bottom-tau Yukawa unification is splitting Higgs masses in such a way that $m_{H_d}>m_{H_u}$ at the GUT scale. In such a case the condition (\ref{mHumHdcond}) may be satisfied already at the GUT scale and the Yukawa-unified solutions with gaugino masses suppressed with respect to the scalar masses become viable, which in turn implies that the SUSY correction (\ref{mbsusycorr}) to the bottom mass may have the magnitude appropriate for top-bottom-tau Yukawa unification. However, since $H_u$ and $H_d$ are part of the same SO(10) multiplet Higgs masses cannot be split without explicit breaking of the SO(10) gauge symmetry. 

Nevertheless, in the effective theory below the GUT scale there is generically additional source of soft scalar masses which is a consequence of spontaneous SO(10) gauge symmetry breakdown. It was shown in Ref.~\refcite{Dterm} that the squared masses of the MSSM scalars acquire additional contribution proportional to their charges under the spontaneously broken U(1) (which is part of SO(10)). The magnitude of this contribution is set by the vacuum expectation value of the $D$-term of the broken U(1) which is model dependent but typically of the order of the soft scalar masses. After taking this effect into account the generic pattern of soft scalar masses in SO(10) models is:
\begin{align}
\label{scalarDterm}
&m_{H_d}^2=m_{10}^2+2D\,, \nn
&m_{H_u}^2=m_{10}^2-2D\,, \nn
&m_{Q,U,E}^2=m_{16}^2+D\,, \nn
&m_{D,L}^2=m_{16}^2-3D\,, \nn
&m_{\nu_R}^2=m_{16}^2+5D\,,
\end{align} 
where $D$ parameterizes the size of a $D$-term contribution. It was shown in Ref.~\refcite{Murayama} that the above structure of soft scalar masses can lead to top-bottom-tau Yukawa unification compatible with REWSB if $D$ is positive and $m_{10}>m_{16}$.

Yukawa unification is also very sensitive to the sign of Higgs-mixing parameter $\mu$. The sign of the dominant SUSY correction to the bottom mass, namely the gluino-sbottom correction, is the same as the sign of $\mu$, as seen from Eq.~(\ref{mbsusycorr}). Therefore, Yukawa unification prefers the negative sign of $\mu$. Nevertheless, top-bottom-tau Yukawa unification can be obtained also for positive $\mu$. The phenomenological implications of Yukawa-unified models differ significantly depending on the sign of $\mu$ so we discuss these two classes of models separately in the following sections.

\section{Positive $\mu$}
\label{sec:posmu}

Top-bottom-tau Yukawa unification in the SO(10) models with positive $\mu$, the universal gaugino masses, $M_{1/2}$, and the universal trilinear couplings, $A_0$, has been extensively studied in the last decade.\cite{Blazek1,Blazek2,Auto,BaerDM} 
For positive $\mu$ the gluino-sbottom correction to the bottom mass has incorrect sign for Yukawa unification so it has to be significantly suppressed, while the chargino-stop correction should be negative implying $A_t<0$. Moreover, the magnitude of the chargino-stop correction has to be large enough to compensate positive gluino-sbottom correction and account for the total finite SUSY correction to the bottom mass of order 10 to 20 percent. From the inspection of Eq.~(\ref{mbsusycorr}) it can be seen that the suppression of the gluino-sbottom correction requires the heavier sbottom to be much heavier than the gluino. However, heavy sbottoms are usually accompanied by heavy stops which tend to suppress the chargino-stop correction. So, in order to obtain the desired magnitude of the finite SUSY correction to the bottom mass very large $|\mu A_t|$ is necessary. This is possible only for a very large  $A_0$ because the RGEs predict $A_t\sim{\mathcal O}(-1.5)M_{1/2}+{\mathcal O}(0.3)A_0$ at the EW scale\cite{Carena} and the gluino should be light to suppress the gluino-sbottom correction. 

Large negative values of the $A$-terms at the GUT scale (and during entire RG evolution down to the EW scale) give very large negative contribution to the mass of the third generation sfermions. Such contribution is especially dangerous for the right-handed sbottom because it is lighter than other squarks already at tree-level (due to the $D$-term contribution) so the RG effects of large $A$-terms can make the right-handed sbottom tachyonic. This is the main reason why the $D$-term splitting of scalar masses leads to rather poor Yukawa unification - worse than at least 8\%.\cite{Baerjustso} \footnote{ Compatibility of the $D$-term splitting of scalar masses, which certainly has much better theoretical justification, with top-bottom-tau Yukawa unification can be improved if the scale of the right-handed neutrinos is relatively small (of order $10^{14}$ GeV) and the soft masses of the third generation sfermions at the GUT scale are allowed to be smaller than the corresponding masses of the first and the second generation. In such a case Yukawa couplings of top, bottom and tau can unify at the GUT scale at the level of 2.5\%.\cite{Baerjustso} } On the other hand, when only the Higgs masses are split while the sfermion masses are universal perfect agreement with top-bottom-tau Yukawa unification can be obtained.\cite{Blazek1,Blazek2,Auto,BaerDM} Such a scenario was dubbed in the literature 'just so' Higgs splitting (HS model).  Even though, the HS model explicitly breaks SO(10) gauge symmetry it gained a lot of attention due to its ability to satisfy very restrictive condition of top-bottom-tau Yukawa unification. 

Early studies of the HS model have proven the existence of the Yukawa-unified solutions with very light gluinos $\sim300-500$ GeV which were within the reach of the LHC at its early stage.\cite{Baer_YukLHC,Baer_YukLHC2} However, gluinos have not been found up to now and the lower limit on the gluino mass in the HS model was found to be 620 GeV at 95\% C.L..\cite{atlas_bjets} Nevertheless, the HS model is not yet ruled out because recently new Yukawa-unified solutions were found which are characterized by heavier gluinos with the mass up to about 1.5 TeV (3 TeV if Yukawa unification at the 10\% level is allowed).\cite{Baer_heaviergluino} Gluino pair production is a main signature of the HS model at the LHC.

It is interesting to note that Yukawa-unified solutions of the HS model are an example of radiatively-driven inverted scalar mass hierarchy which is obtained for $A_0\sim-2m_{16}$ and $M_{1/2}\ll m_{16}$, as shown in Ref.~\refcite{RIMH}. Sfermions of the first two generations have masses $\sim m_{16}\sim{\mathcal O}(20)$ TeV, while the third generation sfermions are much lighter, with the stop masses in the few TeV range. One of the reasons why $m_{16}$ has to be very large are the constraints from rare $B$ decays.\cite{FCNCYukawa} The presence of multi-TeV scalars of the first two generations implies negligible SUSY contribution to $(g-2)_{\mu}$ resulting in more than 3$\sigma$ discrepancy between the theoretical prediction and the experimental result.\cite{g2}   
Since the stops are rather heavy and $|A_t|$ is large (leading to rather large stop-mixing) the lightest Higgs is rather heavy. Therefore, a requirement of the Higgs mass in the vicinity of 125 GeV does not have a large impact on allowed parameter space of the HS model.

Another interesting feature of the MSSM spectra in the HS model is that gaugino masses do not respect the usual GUT relation, $M_1:M_2:M_3\approx 1:2:6$.  
The unusual gaugino mass relation at the EW scale follows from the two-loop effects in the RGEs for gaugino masses which are non-negligible due to the very large values of $A_0$  required for Yukawa coupling unification. One important consequence of the unusual gaugino mass relation is that $M_{1/2}$ may be below 100 GeV without conflict with the lower limit on the chargino mass from LEP. 

In spite of the unusual gaugino mass relation at the EW scale bino is the LSP with the relic abundance few orders of magnitude above the upper WMAP bound. Several resolution to this problem have been proposed.\cite{AutoDM,BaerDM,Baer_axinoDM} One interesting possibility is to introduce non-universal gaugino masses and increase a value of $M_1$ at the GUT scale (keeping $M_2$ and $M_3$ fixed) in such a way that $M_1\approx M_2$ at the EW scale and the bino-wino coannihilations reduce the relic abundance to phenomenologically acceptable values.\cite{AutoDM} The relic abundance of the lightest neutralino can be also reduced if one makes the additional assumption that the strong CP problem is solved using Peccei-Queen mechanism\cite{PQ} which leads to the existence of a pseudoscalar particle, axion,\cite{axion1,axion2} and its supersymmetric partner, axino.\cite{axino} Axino mass, $m_{\tilde{a}}$, may be much smaller than the lightest neutralino mass.\cite{axinoDM1,axinoDM2} In such a case the axino would be the LSP with the relic abundance reduced by a factor of $m_{\tilde{\chi}_1^0}/m_{\tilde{a}}$ with respect to the neutralino LSP case.

\section{Negative $\mu$}
\label{sec:negmu}

Let us now discuss the case of negative $\mu$ in which the threshold correction to the bottom mass is typically negative, as preferred by the Yukawa unification. If the gaugino masses are universal the negative sign of $\mu$ leads to the discrepancy with the experimental results for $(g-2)_{\mu}$ and BR$(b\to s \gamma)$.\cite{Baernegmu1,Baernegmu2} Consistency with these constraints requires $\mu M_2>0$, see e.g. Ref.~\refcite{ChNath}. An SO(10) model with negative $\mu$ satisfying the condition $\mu M_2>0$ was recently proposed in Ref.~\refcite{bop}.\footnote{Top-bottom-tau Yukawa unification with $\mu<0$ and $M_2<0$ was also considered in the context of supersymmetric $SU(4)_c\times SU(2)_L\times SU(2)_R$.\cite{GoKhRaSh,yukuni_su4_125}} In this model the gaugino masses are assumed to be generated by a non-zero $F$-term (which is a SM singlet) in a 24-dimensional representation of ${\rm SU(5)}\subset {\rm SO(10)}$. This leads to the following pattern of the gaugino masses at the GUT scale:\cite{Martin}
\begin{equation}
\label{gauginoratio}
 (M_1,M_2,M_3) = \left(-\frac{1}{2},-\frac{3}{2},1\right)M_{1/2} \,.
\end{equation}
Notice that in spite of the introduction of non-universal gaugino masses only one free parameter, $M_{1/2}$, governs the gaugino sector. In addition, it is assumed that the soft trilinear couplings have a universal value $A_0$ at the GUT scale.\footnote{ This assumption requires the existence of a singlet F-term which dominates the soft trilinear couplings but gives a subdominant contribution to the gaugino masses. An impact of non-universal $A$-terms on the Yukawa unification was investigated in Ref.~\refcite{nonuniA}.} With the above assumptions precise top-bottom-tau Yukawa unification is possible with the $D$-term splitting of scalar masses (\ref{scalarDterm}) for a very wide range of $M_{1/2}$ and $m_{16}$.\footnote{Top-bottom-tau Yukawa unification in a model with $\mu<0$ and non-universal gaugino masses given by Eq.~(\ref{gauginoratio})  can be also obtained for universal sfermion masses and ``ad hoc'' Higgs mass splitting.\cite{Gogoladze_HS_negmu}} Moreover, in some part of parameter space $(g-2)_{\mu}$ and BR$(b\to s \gamma)$ are simultaneously satisfied and the correct relic abundance of the LSP is predicted. The corresponding sparticle spectrum is relatively light. The LHC constraints on such a scenario were investigated in Ref.~\refcite{mbks}.

Yukawa-unified solutions consistent with $(g-2)_{\mu}$ and BR$(b\to s \gamma)$ identified in Ref.~\refcite{bop} are characterized by the Higgs boson mass below about 114.5 GeV. Even after taking into account the fact that the theoretical uncertainty in the prediction of the Higgs mass is about 3 GeV,\cite{Allanach_higgs} such solutions are excluded at 95\% C.L. by the recent ATLAS Higgs search.\cite{atlas_higgs} Therefore, the $(g-2)_{\mu}$ anomaly cannot be explained within the SO(10) model.

Both ATLAS\cite{atlas_higgs} and CMS\cite{cms_higgs} observe an excess of events which may be due to the SM-like Higgs boson with mass of about 125 GeV. In the following we investigate the implications of the 125 GeV Higgs on the SO(10) model. More precisely, due to the theoretical uncertainties mentioned above we focus on the Higgs mass between 122 and 128 GeV. We use SOFTSUSY\cite{softsusy} to solve the 2-loop renormalization group equations and calculate the MSSM spectrum. We neglect the RG effects of the right-handed neutrinos which are expected to be small, especially if constraints from lepton flavour violating processes are taken into account.\cite{neutrinoLFV,neutrinoLFV2}
We quantify the goodness of Yukawa unification using the following quantity:
\begin{equation}
 R\equiv\left.\frac{\max\left(h_t,h_b,h_{\tau}\right)}
{\min\left(h_t,h_b,h_{\tau}\right)}\right|_{\rm GUT} \,.
\end{equation}
We perform a numerical scan of parameter space using Markov Chain Monte Carlo (MCMC) sampling. We apply Metropolis-Hastings algorithm\cite{metropolis1,metropolis2} following Ref.~\refcite{BaerDM}. For every randomly generated point we demand proper REWSB and the neutralino being LSP. We also compute BR$(b\to s\gamma)$, BR$(B_s\to\mu^+\mu^-)$ and the relic abundance of the LSP using appropriate routines in MicrOmegas\cite{Micromega} and apply the following experimental constraints:\cite{bsg,Bsmumu_LHCb,WMAP7}
\begin{align}
 & 2.89\cdot10^{-4}<{\rm BR}(b\to s\gamma)<4.21\cdot10^{-4} \\ 
 & {\rm BR}(B_s\to\mu^+\mu^-)<4.5\cdot10^{-9} \\ 
 &  0.0952<\Omega_{\rm DM}h^2<0.1288   
\end{align}
We found that the lower mass limits on the MSSM particles from the direct accelerator searches do not impose additional constraints on a part of the model parameter space with the lightest Higgs mass in the range under consideration.

\begin{figure}[t]
\centerline{\psfig{file=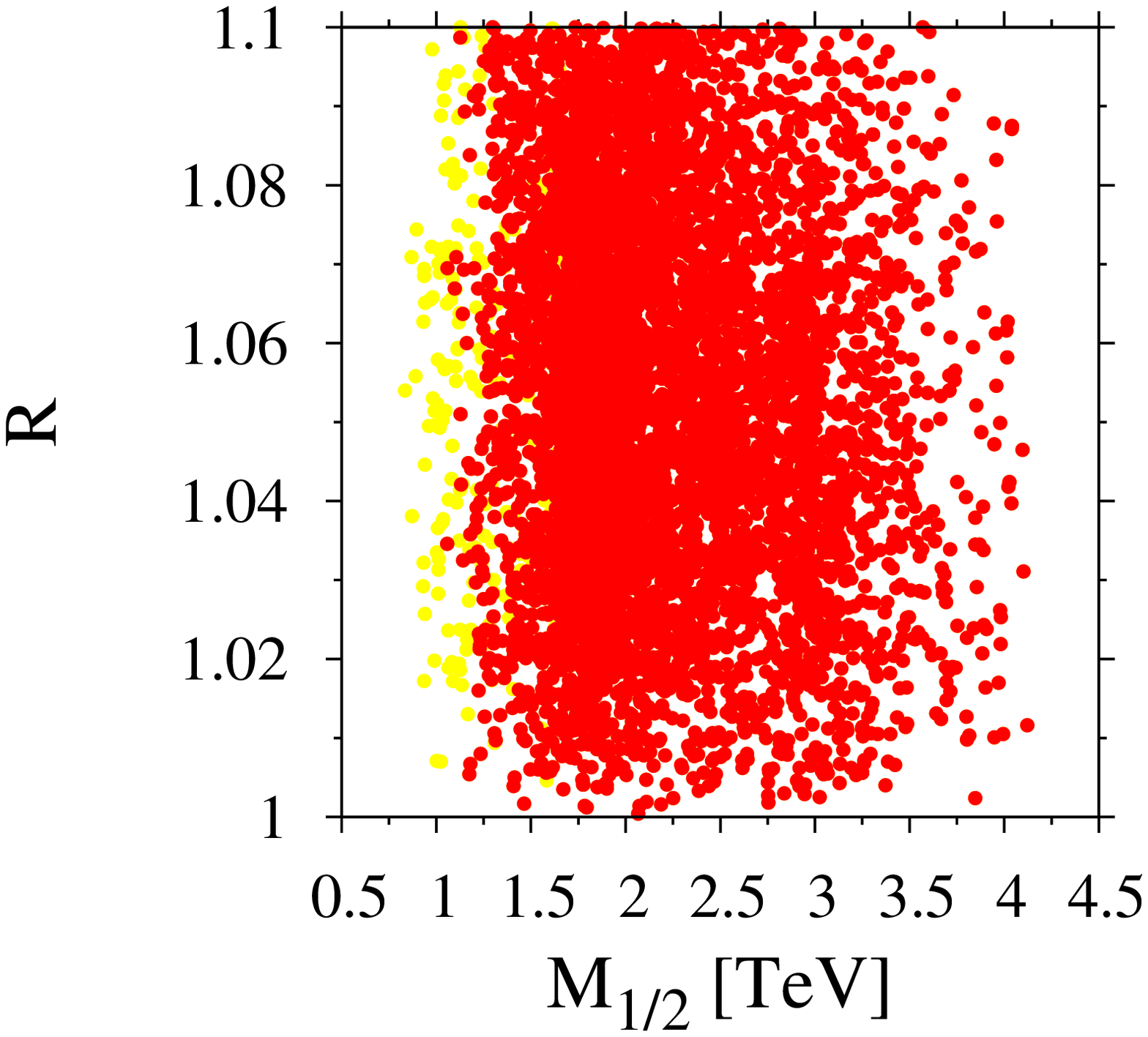,width=0.5\textwidth}
\psfig{file=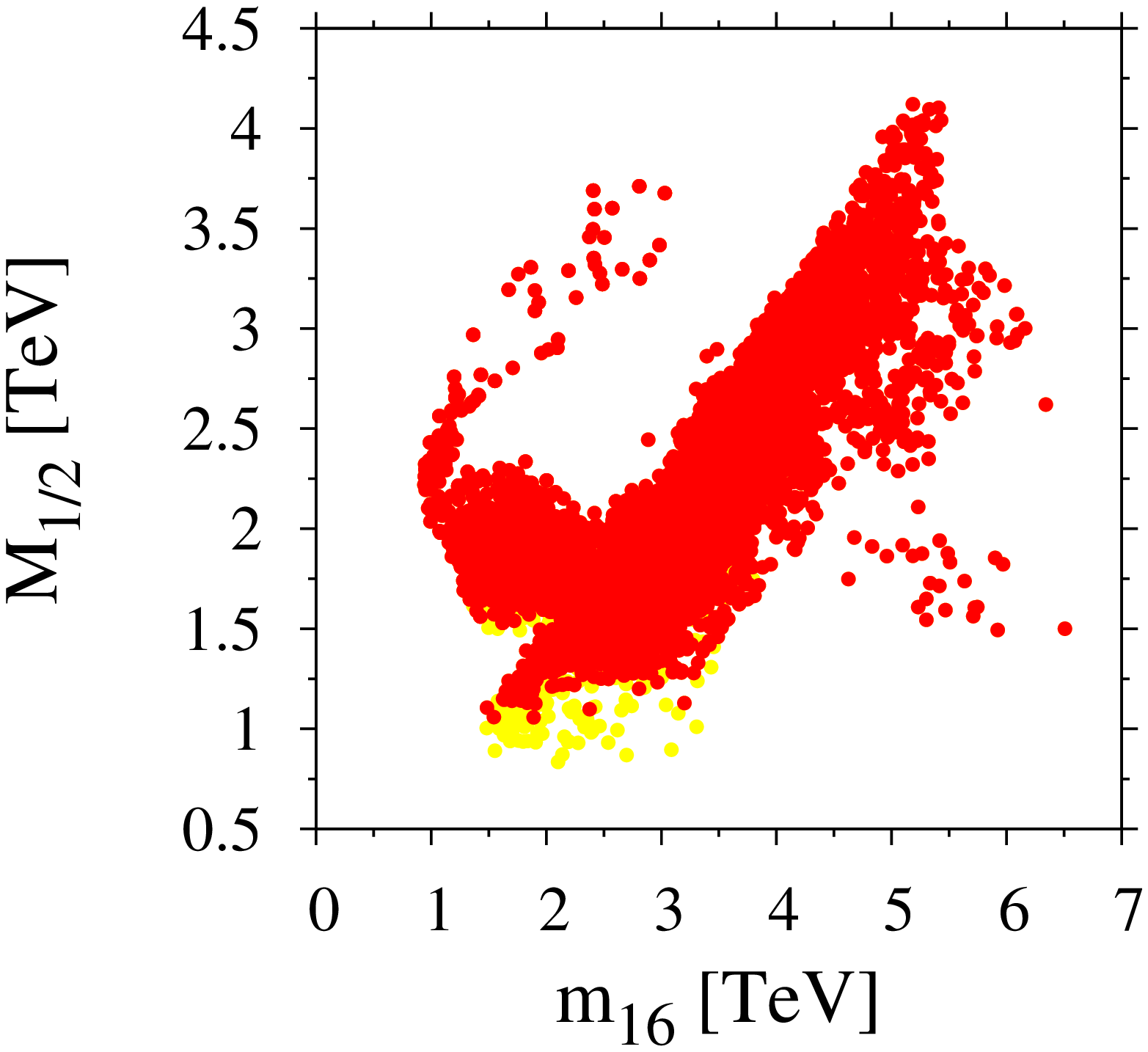,width=0.5\textwidth}
}
\vspace*{8pt}
\caption{Plots in the $R-M_{1/2}$ and $M_{1/2}-m_{16}$ planes. All the points correspond to the Higgs mass in the range 122-128 GeV. The yellow points are consistent with REWSB and neutralino LSP and satisfy particle mass bounds, the constraint from BR$(B_s\to\mu^+\mu^-)$ and the WMAP bounds on the relic abundance of dark matter. The red points satisfy also the constraint from BR$(b\to s\gamma)$. In the $M_{1/2}-m_{16}$ panel all the points have Yukawa unification at the level of $10\%$ or better (i.e. $R\leq1.1$). \protect\label{fig:Rm12_m16}}
\end{figure}

\begin{figure}[t]
\centerline{\psfig{file=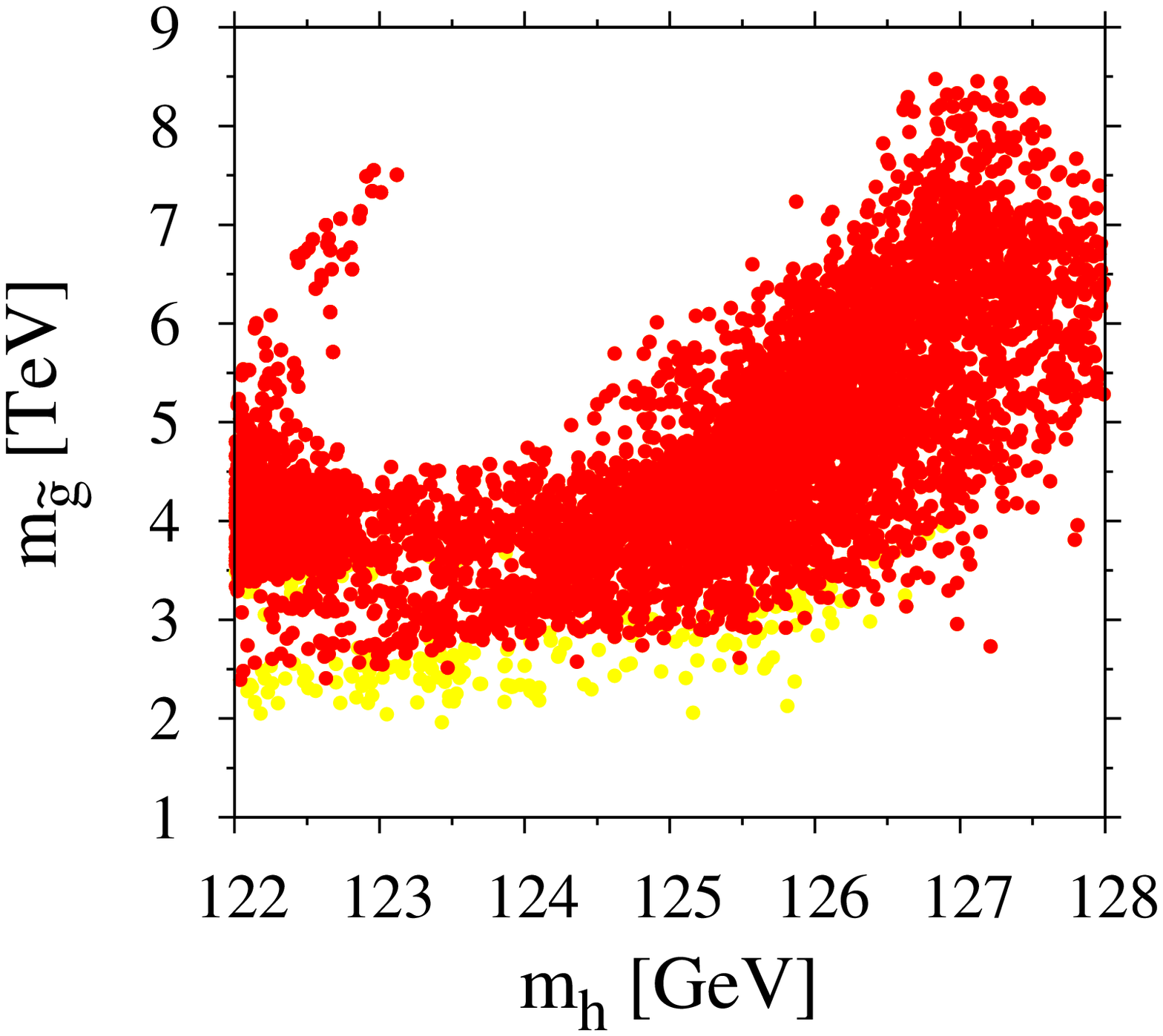,width=0.5\textwidth}
\psfig{file=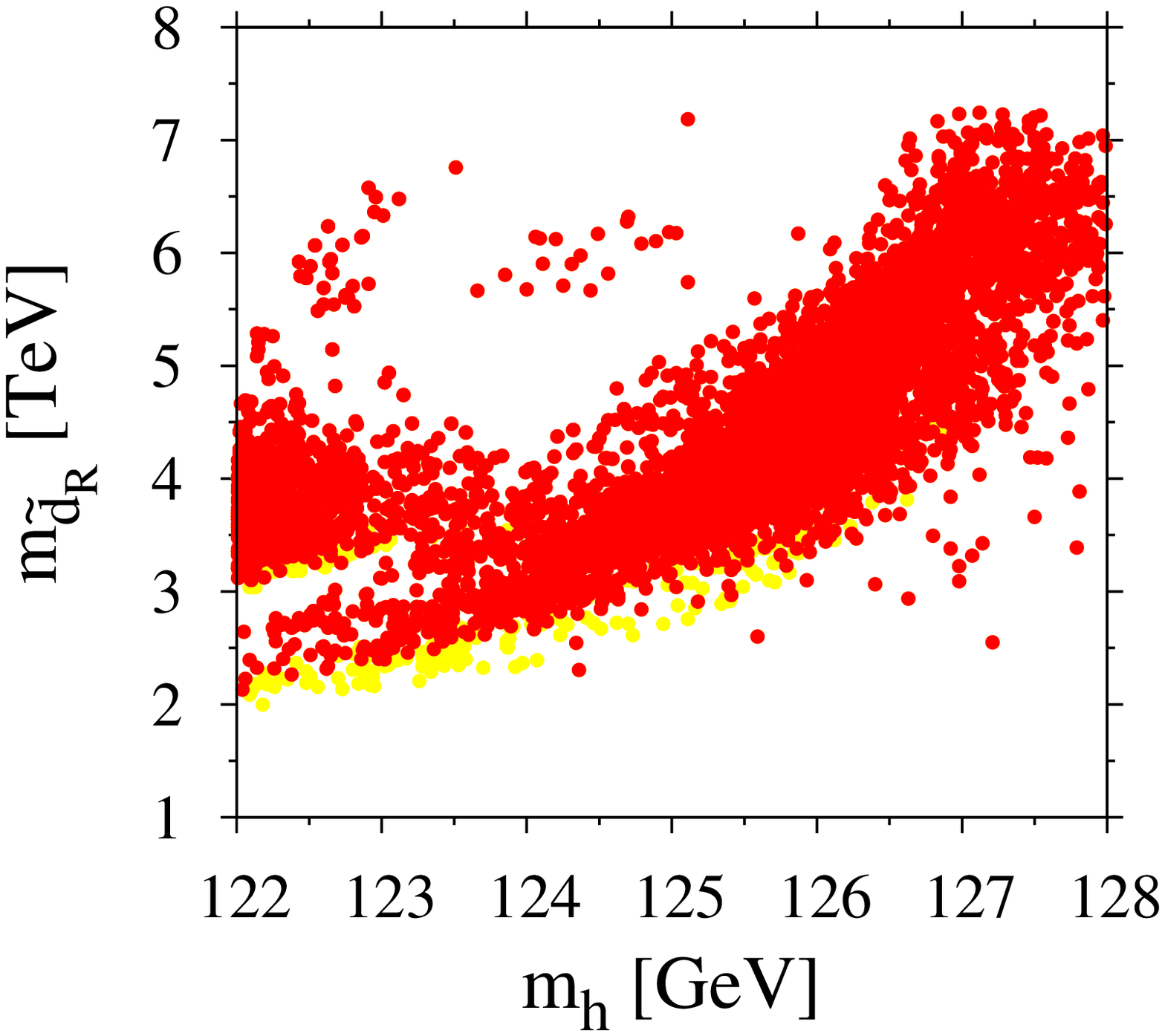,width=0.5\textwidth}
}
\centerline{\psfig{file=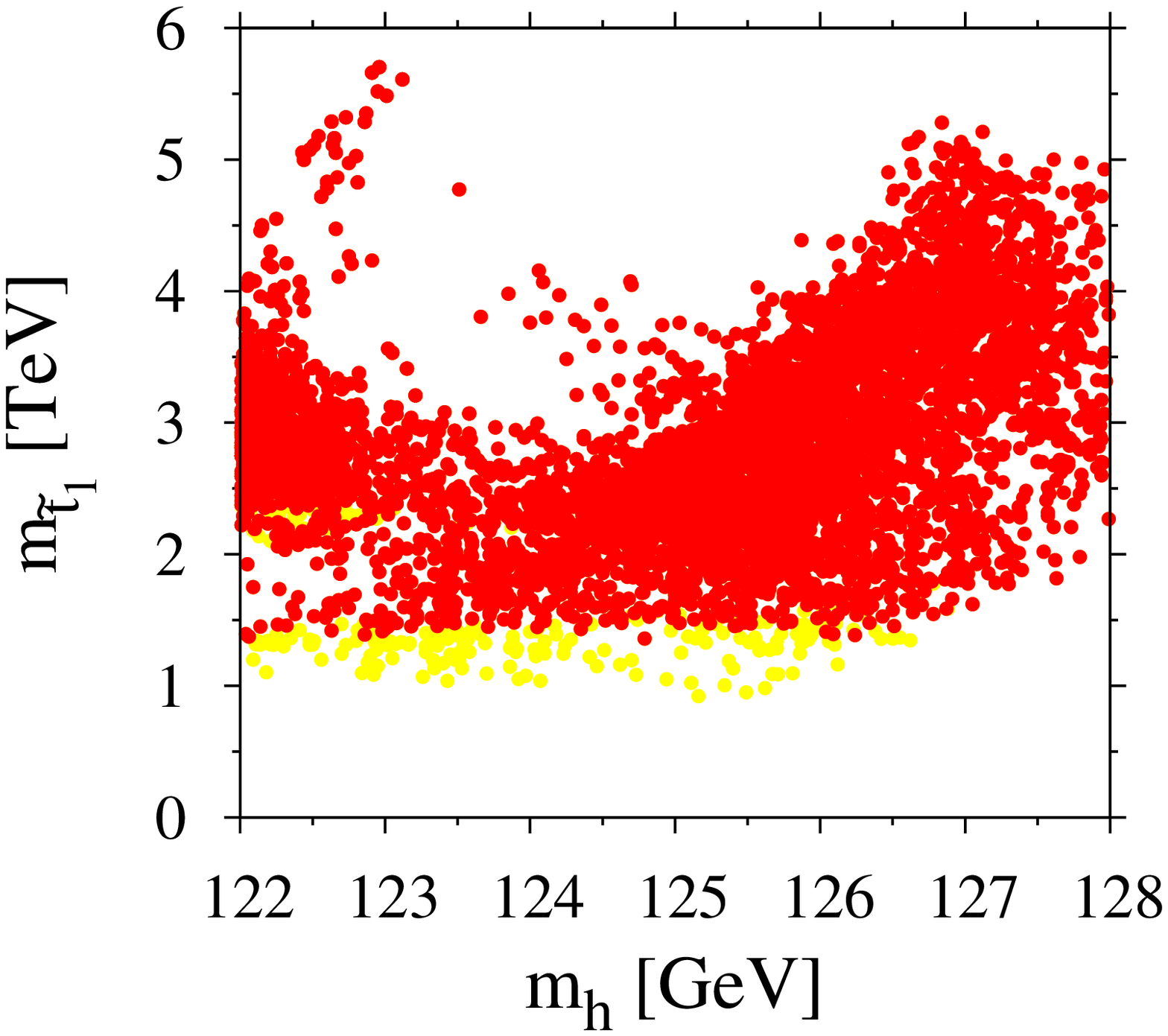,width=0.5\textwidth}
\psfig{file=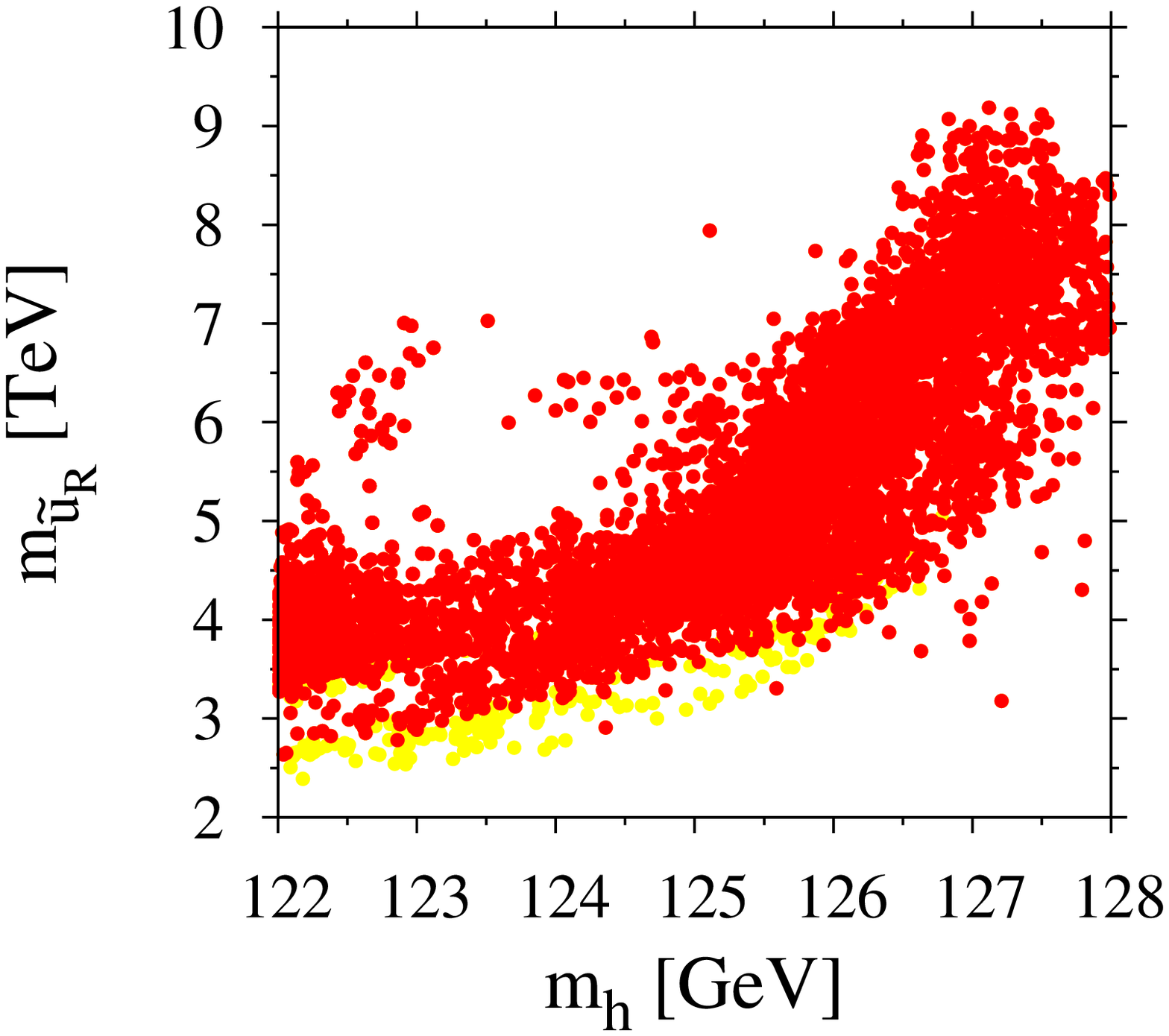,width=0.5\textwidth}
}
\vspace*{8pt}
\caption{Plots in the $m_{\tilde{g}}-m_h$,  $m_{\tilde{d}_R}-m_h$,  $m_{\tilde{t}_1}-m_h$,  $m_{\tilde{u}_R}-m_h$ planes. The color coding is the same as in Figure \ref{fig:Rm12_m16}. All the points have $R\leq1.1$. \protect\label{fig:vs_mh}}
\end{figure}

\begin{figure}[t]
\centerline{\psfig{file=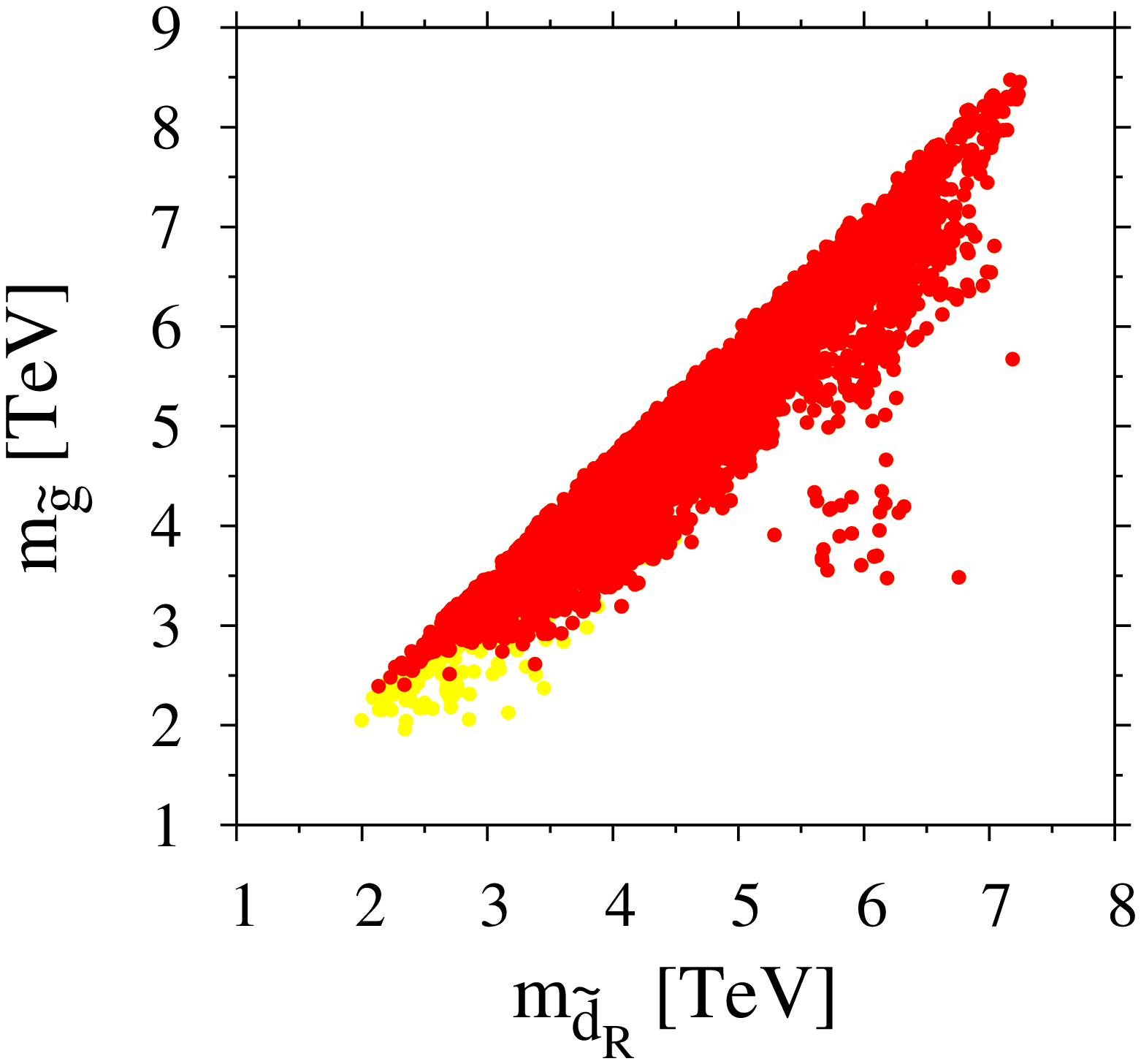,width=0.5\textwidth}
\psfig{file=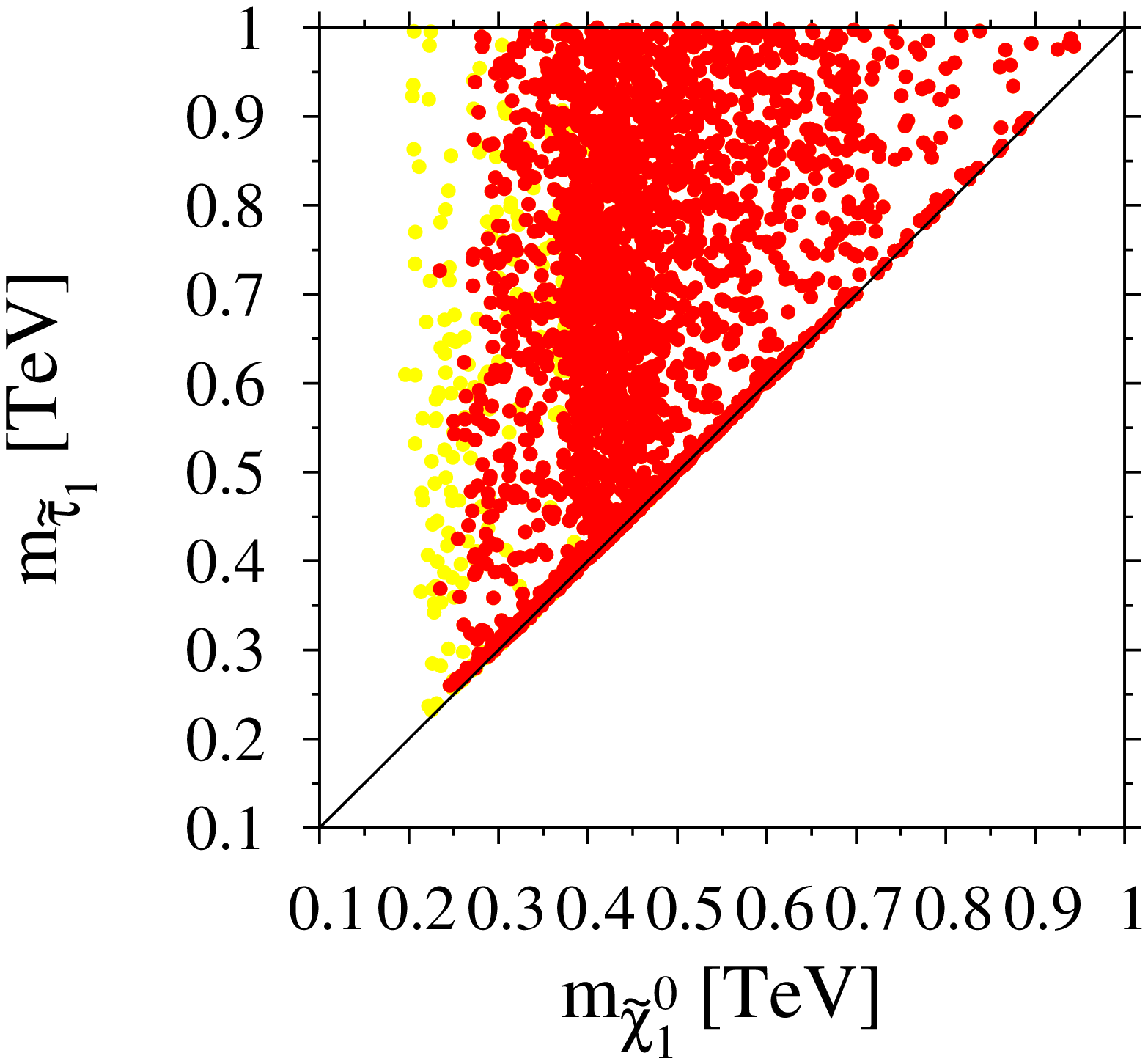,width=0.5\textwidth}
}
\centerline{\psfig{file=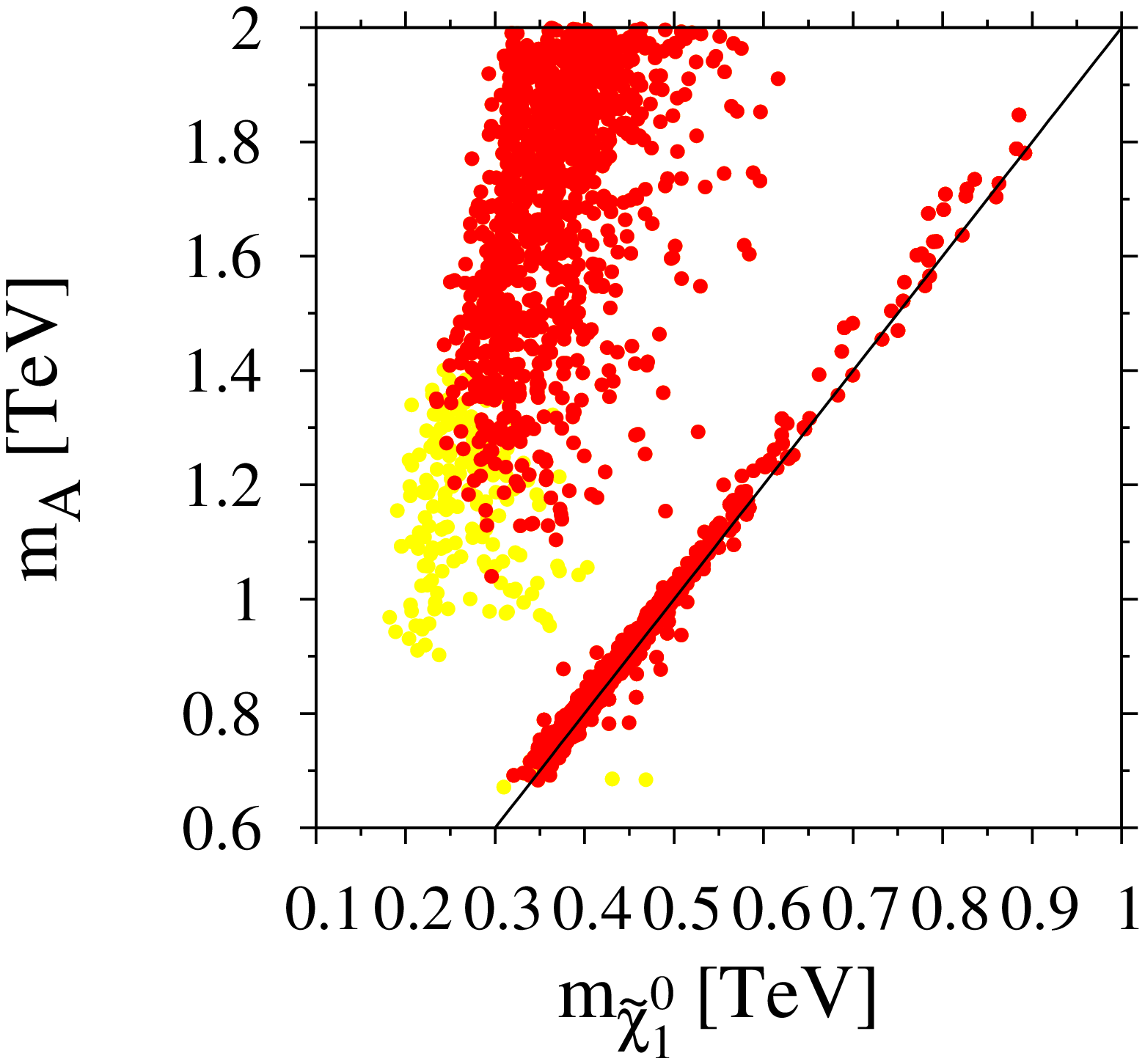,width=0.5\textwidth}
\psfig{file=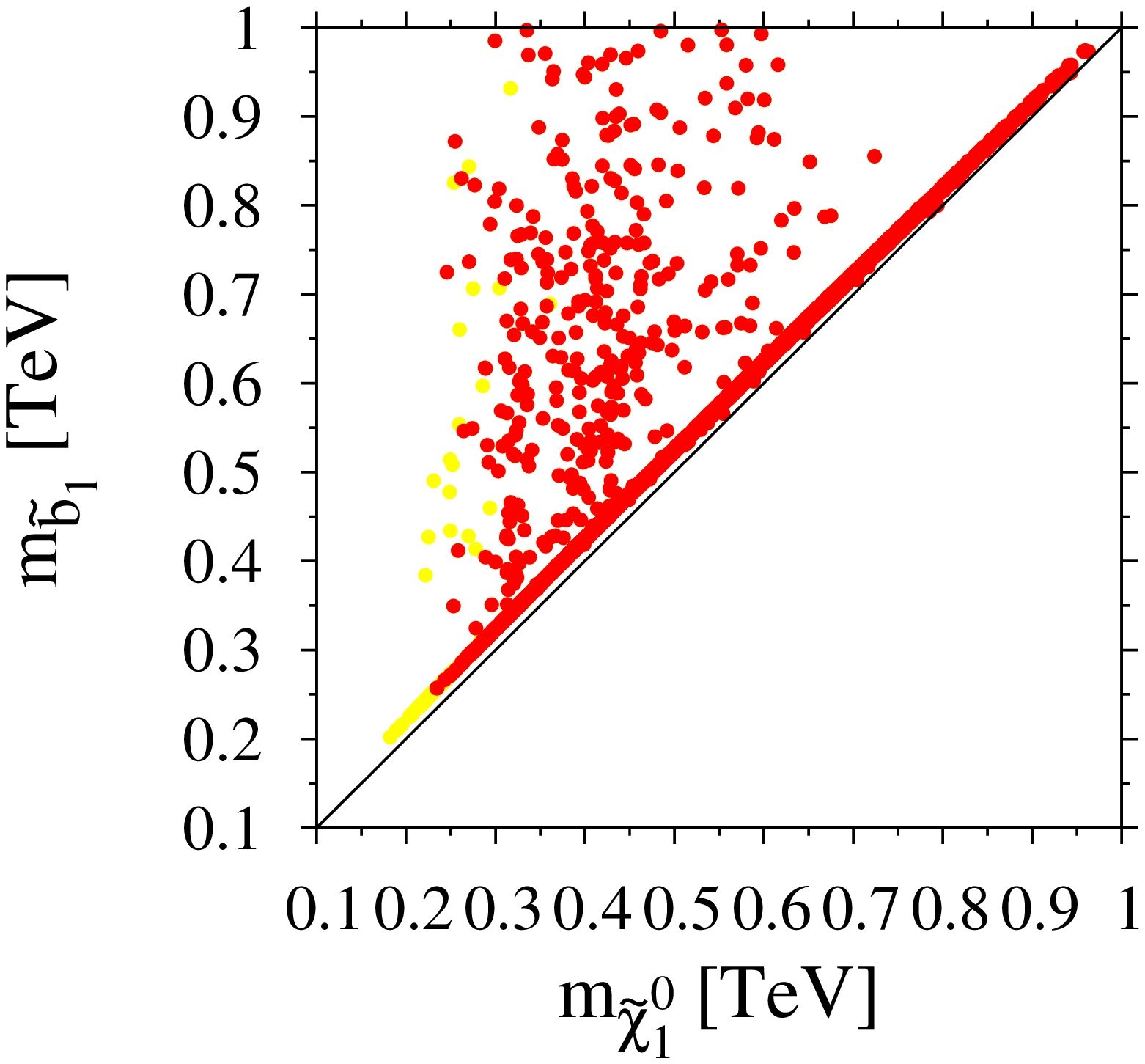,width=0.5\textwidth}
}
\vspace*{8pt}
\caption{Plots in the $m_{\tilde{g}}-m_{\tilde{d}_R}$,  $m_{\tilde{\tau}_1}-m_{\tilde{\chi}_1^0}$,  $m_A-m_{\tilde{\chi}_1^0}$,  $m_{\tilde{b}_1}-m_{\tilde{\chi}_1^0}$ planes. The color coding is the same as in Figure \ref{fig:Rm12_m16}. All the points have $R\leq1.1$ and the Higgs mass in the range 122-128 GeV. \protect\label{fig:vs_lsp}}
\end{figure}

The results of our scan are presented in Figures \ref{fig:Rm12_m16}, \ref{fig:vs_mh} and \ref{fig:vs_lsp}.
Figure \ref{fig:Rm12_m16} demonstrates that very precise top-bottom-tau Yukawa unification (corresponding to $R\approx1$) is compatible with the Higgs mass in the range 122-128 GeV. However, this imposes a lower bound on $M_{1/2}$ and $m_{16}$ of about 1 TeV. If the $b\to s \gamma$ constraint is relaxed slightly smaller values of $M_{1/2}$ are possible.

Large gaugino and sfermion masses at the GUT scale result in the low energy spectrum characterized by rather heavy gluino and the first two generations of sfermions.  Gluino mass is found to be above about 2.5 TeV (2 TeV if the $b\to s \gamma$ constraint is relaxed), as seen from the top left panel of Figure \ref{fig:vs_mh}. The right handed down squark (which is always the lightest first generation squark due to the D-term splitting of scalar masses) has to be heavier than about 2 TeV, while the right handed up squark mass is above 2.5 TeV, as seen from the top right and the bottom right panel of Figure \ref{fig:vs_mh}, respectively. It is also interesting to note that the gluino mass and the right handed down squark mass are positively correlated so the points with the lightest gluino have also the lightest down squark, as seen from the top left panel of Figure \ref{fig:vs_lsp}.

The first two generation sleptons masses are also above 1 TeV. The SUSY contribution to $(g-2)_{\mu}$ is always positive so the discrepancy between  the theory and the experiment is smaller than in the SM. However, we found that the SUSY contribution to $(g-2)_{\mu}$ in Yukawa-unified solutions consistent with the 125 GeV Higgs is at most $4\times10^{-10}$. Therefore, the theoretical prediction for $(g-2)_{\mu}$ is about $3\sigma$ below the experimental central value.

The value of the Higgs boson mass depends crucially on the stop sector. In general MSSM, the 125 GeV Higgs requires large loop contribution from stop-mixing and the average stop masses of about 1 TeV. It can be seen from the bottom left panel of Figure \ref{fig:vs_mh} that Yukawa-unified solutions consistent with the 125 GeV Higgs are characterized by the lighter stop as light as 900 GeV. Since the stop contribution to BR$(b\to s\gamma)$ is the largest for the maximal stop mixing, the lighter stop mass has to be above about 1.4 TeV if the $b\to s \gamma$ constraint is taken into account. 

The Yukawa-unified solutions with the lighter stop are characterized by large negative values of $A_0$, of order few TeV. This is a consequence of the fact that the large stop mixing requires large $A$-terms at the GUT scale as long as two loop effects in the RGEs are negligible.\cite{Brummer} It was recently pointed out that in the so-called inverted hierarchy scenario, with the first two generation sfermion masses at the GUT scale much larger than those of the third generation, the maximal stop mixing can be obtained even for vanishing $A$-terms due to large two-loop effects from the heavy first two generations.\cite{IHmaxmix} However, the search for the Yukawa-unified solutions with small $A_0$ in the inverted hierarchy scenario is beyond the scope of the present paper.

The lightest supersymmetric particle is a bino-like neutralino. The LSP can play a role of dark matter in the Universe in three regions of parameter space. It can be seen from the top right panel of Figure \ref{fig:vs_lsp} that the LSP with mass between about 250 and 900 GeV can be strongly degenerate with the stau resulting in very efficient stau coannihilations. The corresponding benchmark point is given as a Point A in Table \ref{tab:benchmarks}. Notice that apart from the stau also the lighter sbottom is relatively light. The latter should be within the reach of the LHC. 

\begin{table}[t!]
\tbl{Several benchmark points with top-bottom-tau Yukawa unification characterized by the Higgs mass close to 125 GeV. All dimensionful parameters are in GeV.  Point A represents the stau-coannihilation scenario, point B resonant annihilations via pseudoscalar Higgs exchange, while point C is an example of   small mass splitting between the bino LSP and the lighter sbottom. }
{
\begin{tabular}{|c|ccc|}
\hline
\hline
                 & Point A & Point B & Point C  \\
\hline
$m_{16}$        & 2041  & 5710  & 2326  \\
$M_{1/2}$        & 1303 & 1563  & 1307   \\
$m_{10}/m_{16}$   & 1.672 & 1.093  & 1.596   \\
$D/m_{16}^2$         & 0.1771 & 0.042  & 0.1557   \\
$A_0/m_{16}$      & -2.987  &  -1.277  &  -2.765   \\
$\tan\beta$      & 44.37 & 47.91 & 44.99    \\

\hline
$\mu$            & -1207  & -1782  & -1170  \\

\hline
$m_h$            & 124.4  & 124.4 & 124.6   \\
$m_H$            & 1299  & 704   & 1303   \\
$m_A$            & 1299 & 705  & 1303   \\
$m_{H^{\pm}}$    & 1302  & 710    & 1305   \\

\hline
$m_{\tilde{\chi}^0_{1,2}}$
                 & 291, 1195 & 357, 1781  & 292, 1161   \\
$m_{\tilde{\chi}^0_{3,4}}$
                 & 1195, 1647 & 1795, 2028  & 1167, 1653    \\

$m_{\tilde{\chi}^{\pm}_{1,2}}$
                 & 1195, 1647 & 1781, 2028  & 1161, 1653   \\
$m_{\tilde{g}}$  & 2908 & 3607  & 2932   \\

\hline $m_{ \tilde{u}_{L,R}}$
                 & 3499, 3264  & 6573, 6402 & 3687, 3462    \\
$m_{\tilde{t}_{1,2}}$
                 & 1431, 2076 & 3733, 4223  & 1478, 2119  \\
\hline $m_{ \tilde{d}_{L,R}}$
                 & 3500, 2804  & 6573, 5979 & 3687, 2966    \\
$m_{\tilde{b}_{1,2}}$
                 & 530, 2054  & 3473, 4214  & 316, 2097   \\
\hline
$m_{\tilde{\nu}_{1,2}}$
                 & 1849  & 5509  & 2082  \\
$m_{\tilde{\nu}_{3}}$
                 & 965 & 4576  & 1168  \\
\hline
$m_{ \tilde{e}_{L,R}}$
                & 1851, 2260  & 5510, 5854  & 2084, 2545   \\
$m_{\tilde{\tau}_{1,2}}$
                & 298, 975 & 3917, 4577  & 691, 1175  \\

\hline

$\Omega_{DM}h^{2}$ &  0.123 & 0.125  & 0.118   \\
BR$(b\to s\gamma)$ &  $4.2\times 10^{-4} $  &  $4\times 10^{-4} $ & $4.2\times 10^{-4} $ 	\\
BR$(B_s\to \mu^+ \mu^-)$ &  $1.5\times 10^{-9} $ & $2.9\times 10^{-9} $ & $1.5\times 10^{-9} $  	\\
$a_{\mu}^{\rm SUSY}$  & $2.2\times 10^{-10} $ & $5\times 10^{-11} $  & $2\times 10^{-10} $   \\
\hline
R & 1.02 & 1.02 & 1.01   \\

\hline
\hline
\end{tabular}
}
\label{tab:benchmarks}
\end{table}

The existence of the A-funnel region, with $m_A\approx2m_{\tilde{\chi}_1^0}$, is presented in the bottom left panel of Figure \ref{fig:vs_lsp}. In this region the LSP mass is between about 300 and 900 GeV. The pseudoscalar Higgs in such a case can be as light as about 700 GeV. This is not much above the present lower limit from CMS, $m_A\gtrsim450$ GeV for $\tan\beta\sim45$, and there are good prospect for a discovery of such a light CP-odd Higgs at the LHC, see e.g. Ref.~\refcite{mA_prospects}. The corresponding benchmark point is given as a point B in Table \ref{tab:benchmarks}. Notice that in spite of a large value of $\tan\beta$ and relatively light CP-odd Higgs the constraint from BR$(B_s\to\mu^+\mu^-)$ is satisfied. This is a consequence of rather heavy stops which suppress the SUSY contribution to BR$(B_s\to\mu^+\mu^-)$.   

Both, the stau-coannihilation and the A-funnel scenarios, were identified in the SO(10) models before. However, we also found a novel scenario leading to correct relic abundance of the LSP due to efficient sbottom coannihilations. In such a case the LSP mass is between about 250 GeV (200 GeV if the $b\to s \gamma$ constraint is relaxed) and 1 TeV while the lightest sbottom is only few tens of GeV heavier than the LSP, as seen from the bottom right panel of Figure \ref{fig:vs_lsp}. The corresponding benchmark point is given as Point C in Table \ref{tab:benchmarks}. Notice that the mass splitting between the sbottom and the LSP required to reduce the relic abundance of the LSP to acceptable values is larger than the mass splitting between the stau and the LSP in the case of stau-coannihilations. We should stress that all the Yukawa-unified solutions with sbottom coannihilations found in our scan are consistent with the recent ATLAS search for direct sbottom pair production\cite{Atlas_directsbottom} since the latter set the lower limit on the sbottom mass only for the LSP mass below 150 GeV.
In general compressed spectra are challenging from the experimental point of view. In the case of the sbottom coannihilation scenario, b-jets from the sbottom decays to LSP are expected to have a small $p_T$ and usually do not pass the selection cuts used in typical SUSY searches. Nevertheless, it was shown in Ref.~\refcite{sbottomNLSP} that the sbottom NLSP with mass $\sim{\mathcal O}(500)$ GeV can be discovered at the LHC with $\sqrt{s}=14$ TeV even if the mass splitting between the sbottom and the LSP is as small as $10\%$.

We should also comment on the fact that top-bottom-tau Yukawa unification can be realized in a model with non-universal gaugino masses given by Eq.~(\ref{gauginoratio}) without the Higgs mass splitting or even for universal scalar masses (i.e. for $D=0$ and $m_{10}=m_{16}$).\cite{Gogoladze_F24_uniscalar} In such a case Yukawa-unified solutions predict too large relic abundance of the LSP if the Higgs mass is about 125 GeV. Moreover, squarks and gluino are heavier as compared to the case with a non-zero $D$-term contribution. Most likely, all the sparticles in such a scenario are outside the LHC reach. On the other hand, the pseudoscalar Higgs is generically light and may be easily discovered at the LHC.\footnote{As a matter of fact, some of the benchmark points presented in Ref.~\refcite{Gogoladze_F24_uniscalar} are already excluded at 95\% C.L. by the CMS search for the pseudoscalar Higgs.\cite{cms_mAbound} }

\section{Summary}
\label{sec:sum}

The lightest Higgs mass in the vicinity of 125 GeV requires typically rather heavy SUSY spectrum in the MSSM. This also applies to the SUSY SO(10) GUT models predicting top-bottom-tau Yukawa unification. Nevertheless, not all the MSSM particles have to be heavy.

Phenomenological implications of the SO(10) models strongly depend on the sign of $\mu$. In the case of positive $\mu$, gluinos are the only colored sparticles within the reach of the LHC. Therefore, the main LHC signature of this class of models is significant gluino pair production. Gluinos from Yukawa-unified models with $\mu>0$ can be light enough to be detected during the present run of the LHC at $\sqrt{s}=8$ TeV.

For testing the SO(10) models with negative $\mu$ one needs to wait a little bit longer but this class of models can be tested at the LHC in several ways. In the A-funnel region, the pseudoscalar Higgs can be as light as 700 GeV. Due to big enhancement of the pseudoscalar Higgs production cross-section by large values of $\tan\beta$, as predicted by top-bottom-tau Yukawa unification, the LHC may have sensitivity to such a light pseudoscalar Higgs even with $\sqrt{s}=8$ TeV. In addition, the lighter sbottom can be very light, with the mass as small as 250 GeV, and strongly degenerate with the bino LSP leading to the prediction of the correct relic abundance of the dark matter due to efficient sbottom coannihilations. There are good prospects for a discovery of such a light sbottom NLSP at the LHC. Gluino and squarks of the first two generations are rather heavy. Nevertheless, gluino and the right-handed down squark with mass 2.5 and 2 TeV, respectively, are possible, which makes them accessible at the LHC with $\sqrt{s}=14$ TeV.

\section*{Acknowledgments}

This work has been partially supported by STFC. MB would like to thank M. Olechowski, S. Pokorski and K. Sakurai for a fruitful collaboration on Yukawa unification and the Cambridge SUSY Working group for helpful discussions.

\end{document}